\documentclass[doublecol]{epl2} 

\usepackage{graphicx}
\usepackage{dcolumn}
\usepackage{bm}
\usepackage{subfigure}
\usepackage{psfrag}
\usepackage{amsmath}
\usepackage{amssymb}

\title{Inertial effects on two-particle relative dispersion in turbulent flows}

\author{Mathieu Gibert\inst{1,4}\thanks{E-mail: \email{mathieu.gibert@ds.mpg.de}} \and Haitao Xu\inst{1,4}\thanks{E-mail: \email{haitao.xu@ds.mpg.de}} \and Eberhard Bodenschatz\inst{1,2,3,4}\thanks{E-mail: \email{eberhard.bodenschatz@ds.mpg.de}}}
\shortauthor{M. Gibert \etal}
\institute{                    
  \inst{1} Max Planck Institute for Dynamics and Self Organization - D-37073 G\"ottingen, Germany\\
  \inst{2} Institute for Nonlinear Dynamics, University of G\"ottingen - D-37073 G\"ottingen, Germany\\
  \inst{3} Laboratory of Atomic and Solid-State Physics and Sibley School of Mechanical and Aerospace Engineering - Cornell University, Ithaca, New York 14853\\
  \inst{4} International Collaboration for Turbulence Research
}

\abstract{
We report experimental results on the relative motion of  pairs of solid spheric particles with initial separations in the inertial range of  fully developed turbulence in water. The particle densities were in the range of $1 \lessapprox \rho_{p}/\rho_{f} \lessapprox 8$, \textit{i.e.}, from neutrally buoyant to highly inertial; and  their sizes were of the Kolmogorov scale.  For all particles, we observed a  Batchelor like regime, in which particles separated ballistically. Similar to the Batchelor regime for tracers, this  regime was observed in the early stages of the relative separation for times $t \lessapprox 0.1 t_0$ with $t_0$ determined by the turbulence energy dissipation rate and the initial separation between particle pairs. In this time interval heavier particles separated faster than fluid tracers. The second order Eulerian velocity structure functions was found to increase with density. In other words, both observations show that  the relative velocity between inertial particles was larger than that between tracers.   Based on the widely used, simplified equation of motion for inertial point-particles, we derived a model that shows an increase in relative velocity between inertial particles. In its scale dependence, however,  it disagrees quantitatively with the experimental results. This we attribute to the preferential sampling of the flow field  by inertial particles, which is not captured by the model.
}
                              
\begin{document}
\maketitle
Many natural and industrial phenomena involve the interaction between turbulent flows and inertial particles, \textit{i.e.}, particles that do not passively follow the fluid motion.
The transport of inertial particles by turbulent flows plays a fundamental role in a vast range of systems, such as, sedimentation in estuaries and rivers \cite{seminara:2010}, the dynamics of plankton in the ocean \cite{Schmitt:2008p2887,denman:1995},  deep sea land slides \cite{Meiburg2009rev}, dust in tornadoes \cite{Lewellen:2008p2927}, sandstorms over deserts or on Mars \cite{Waller:2008p3195}, the dynamics and collisions of water droplets in clouds \cite{Shaw:2003p769} and to the dynamics and clustering of interstellar dust in planet formation \cite{Klahr:2006p3103}.
Quantitative experimental studies on the dynamics of inertial particles in turbulence therefore provide much needed data to verify theoretical models and to test both simulations of idealized particle dynamics in turbulent flows\cite{Bec:2006p656,Ferrante:2003p3145,Bec:submitted}, as well as, coarse grained simulations of natural and technical flows that rely on so called "subgrid" models that parametrize the small, unresolved scales of turbulence \cite{Zaichik:2009p3167}. 

The dynamics of single inertial particles in turbulence has been studied experimentally since the pioneering work by Snyder \& Lumley~\cite{snyder:1971} and has been a topic under extensive experimental~\cite{Ayyalasomayajula:2006p267,Qureshi:2007p1316,Xu:2008p1025,Volk:2008p271}, theoretical~\cite{MaxeyRiley,GATIGNOL:1983p3869} and numerical~\cite{Bec:2006p656,Ferrante:2003p3145} investigations.  Interested readers are referred to a recent review~\cite{PartRev09}.

In this letter, we report an experimental investigation of relative dispersion between two inertial particles  with initial separations in the inertial range of  fully developed turbulent water flow. We focused on solid spherical particles with densities larger than or nearly the same as the fluid. The size of particles were that of the Kolmogorov scale, the smallest scale in turbulence. The particle Stokes number (see definition later) of the particles was in the range between 0.09 and 0.5. 
For these particles, we observed a  Batchelor like regime, in which particles separated ballistically. Similar to the case of fluid tracers~\cite{Bourgoin:2006p391,Ouellette:2006p384}, this regime existed in the early  stage of relative separation for times below a timescale $t_0$ determined by the turbulence energy dissipation rate and the initial separation between particle pairs. In this  Bachelor like regime, particles with larger Stokes numbers separated faster, \textit{i.e.}, the relative velocity between heavy particles were larger than that between fluid tracers at the same separations. This was also supported by the measured second order Eulerian velocity structure functions.  Based on the widely used, simplified equation of motion for inertial point particles ~\cite{MaxeyRiley}, we derived a model that captures the observed increase in relative velocity between inertial particles. In its scale dependence, however,  it disagrees quantitatively with the experimental results. This can most likely be attributed to the preferential sampling of the flow field  by inertial particles, which is not captured by the model. 

\begin{figure}
\begin{center}
  \subfigure[]{\includegraphics[width=4.6cm]{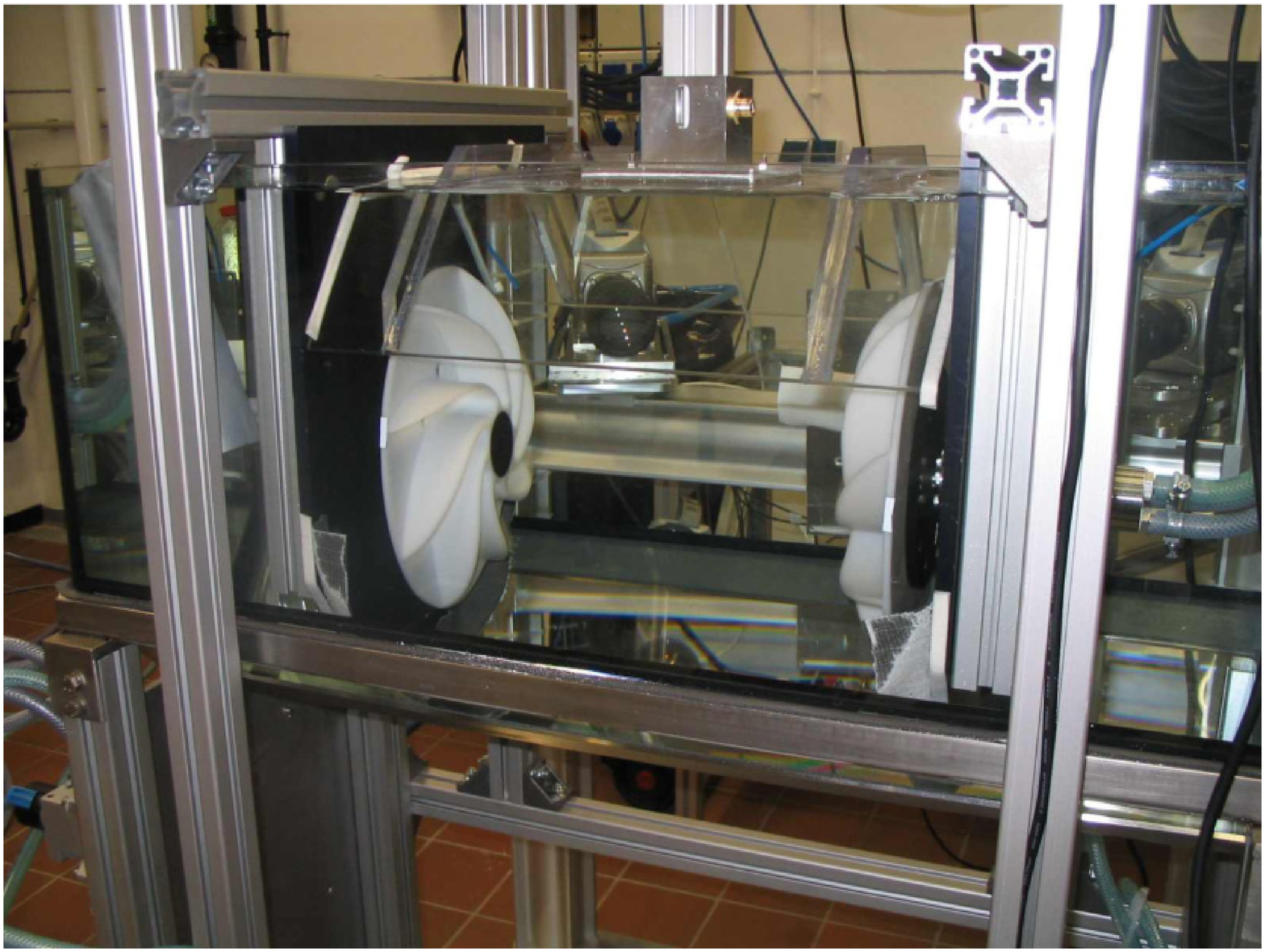}}\quad
  \subfigure[]{\includegraphics[width=3cm]{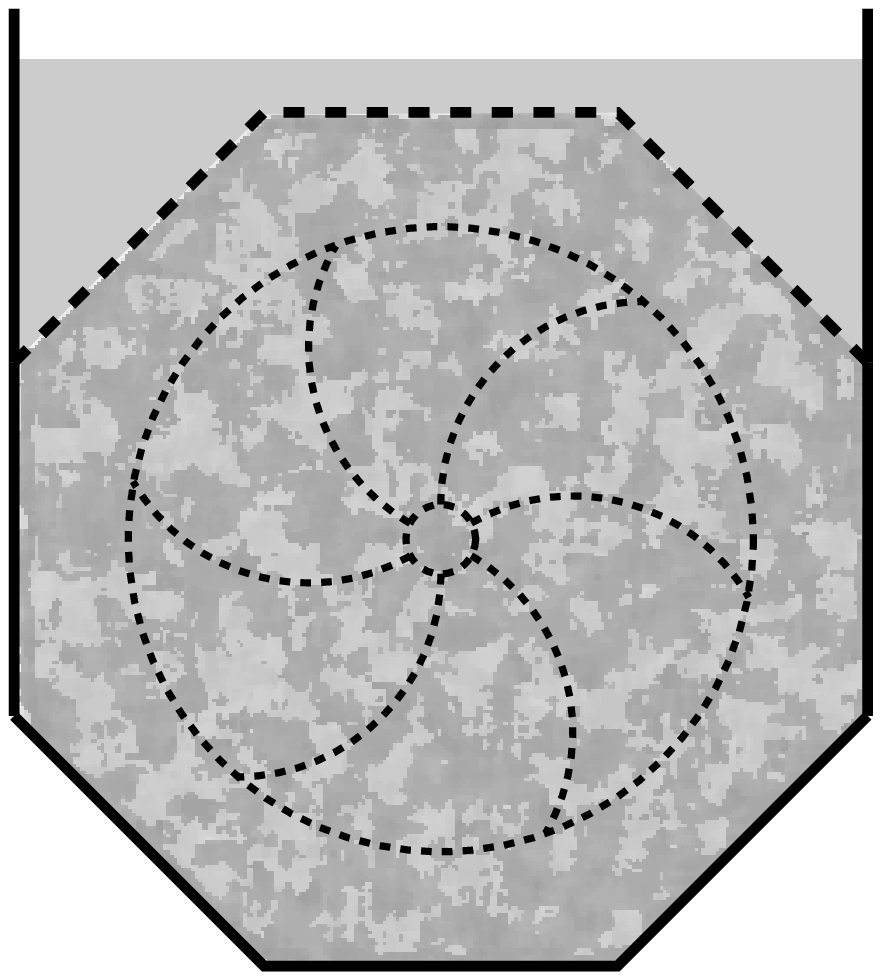}}
\end{center}
\caption{\label{PhotoExp}(a)The experimental setup. (b) Schematic side view of the apparatus. The solid lines represent the walls of the glass container, the dashed lines represent the removable cover. During the experiments, the "aquarium" was filled with water to a level above the cover, as shown by the grey shadow. Two propellers  created a highly turbulent flow in the hexagonal-cylindrical shaped region determined by the cover and the aquarium walls, while the water above the cover remained quiescent.}
\end{figure}

We studied the motion of solid spherical particles in a von K\'arm\'an swirling water flow. The turbulent flow was generated by two counter-rotating propellers submerged in an aquarium and was confined on top by a removable cover with rubber seals along the rim (see Figure~\ref{PhotoExp}). 
Two underwater air-motors drove the baffled propellers so that a highly intense turbulent flow could be produced in an apparatus with moderate size.
The diameter of the propellers was 28 cm. The turbulence chamber, shaped as a hexagonal cylinder, measured $40$cm along the axis of the propellers and $38$cm in both height (vertically) and width (horizontally) in the cross-section. The rotating axis of the propellers was in the horizontal direction so that  the heavy particles settling down towards the bottom of the apparatus were entrained by the strong sweeping of the fluid near the bottom surface. In this way particles stayed suspended  in the measurement volume. Measurements were conducted at the center of the apparatus where the mean fluid velocity was small~\cite{Voth:2002p432} . 

We used three types of particles with the same average size, but different density, as shown in Table~\ref{tab:ExperimentParamParticles}. We adjusted the motors speed such that the Kolmogorov scale of the flow was close to the particle size. In this case, the  almost neutrally buoyant polystyrene particles, behaved as tracer particles, as shown in previous experiments~\cite{Voth:2002p432}. The other two types of particles: the Barium-Titanium glass spheres and the stainless steel spheres, are significantly heavier than water. To characterize particle inertia, we define the Stokes number as $St \equiv  \frac{1}{12\beta}(d_{p}/\eta)^{2}$, where $d_{p}$ is the particle diameter, $\beta\equiv 3\rho_{f}/(2\rho_{p}+\rho_{f})$ is the modified density ratio with $\rho_{p}$ and $\rho_f$ being, respectively, the particle and the fluid densities, and $\eta$ is the Kolmogorov length scale of the turbulence. The Stokes numbers for these particles were $0.09$, $0.27$ and $0.50$, respectively.

\begin{table}
\begin{center}
\begin{tabular}{c | c | c | c | c}
 & $\rho_{p}/\rho_{f}$ & $d_p$ ($\mu$m) & $d_{p}/\eta$ & $St$\\
\hline
\hline
Polystyrene & $1.06$ & $74 \pm 10$ & $0.98$ & $0.09$\\ 
\hline
Glass & $4$ & $75 \pm 8$ & $1.04$ & $0.27$\\ 
\hline
Steel & $7.8$ & $75 \pm 15$ & $1.04$ & $0.50$\\ 
\end{tabular}
\end{center}
\caption{Characteristics of the different particles used in the experiments. The Stokes number is defined as $St \equiv  \frac{1}{12\beta}(d_{p}/\eta)^{2}$, where $d_{p}$ is the particle diameter, $\beta\equiv 3\rho_{f}/(2\rho_{p}+\rho_{f})$ is the modified density ratio with $\rho_{p}$ and $\rho_f$ being, respectively, the particle and the fluid densities, and $\eta$ is the Kolmogorov length scale of the turbulence.}
\label{tab:ExperimentParamParticles}
\end{table}

We measured three-dimensional particle trajectories with high spatial and temporal resolutions using Lagrangian Particle Tracking~\cite{Ouellette:2006p384,Xu:2008p1921} with three high speed CMOS cameras (Phantom V10, manufactured by Vision Research Inc., Wayne, USA). 
The particle velocities and accelerations of the particles were then obtained by smoothing and differentiating the trajectories~\cite{Mordant:2004p245}. 
The measurement volume of LPT was determined by the intersection of the fields of view from all three cameras. To avoid the effect of biased sampling due to the shape of the measurement volume, we used only particle trajectories inside a sphere with diameter of $22$ mm located at the center of the apparatus. We measured $3\times 10^{7}$ data points for the polystyrene particles, $8\times 10^{7}$ for the glass particles and $3\times 10^{6}$ for the steel particles. The LPT provided simultaneous measurements of multiple particles and hence the data could also be used to extract Eulerian statistics even though the LPT technique itself is inherently Lagrangian. For example, knowing velocities of different particles at the same instant, \textit{i.e.}, on the same ``frame'', one could measure velocity increments at distances determined by particle positions. The Eulerian velocity structure functions were then obtained by collecting the statistics of velocity increments over many frames.
The turbulence properties, such as the energy dissipation rate per unit mass $\epsilon$ and the integral scale $L$, were inferred from the Eulerian statistics using polystyrene particles as fluid tracers.
In the inertial range $\eta \ll r \ll L$ of homogeneous and isotropic turbulence, neglecting intermittency corrections, the second order longitudinal and transverse velocity structure functions should scale as
\begin{equation}
D_{LL}(r) = C_2 (\epsilon r)^{2/3},
\end{equation}
and
\begin{equation}
D_{NN}(r) = \frac{4}{3}C_2 (\epsilon r)^{2/3},
\end{equation}
where $C_2$ is expected to be a universal constant. We used $C_2 = 2.1$ as suggested from a compilation of available data~\cite{sreenivasan:1995}.
In addition, there were two exact inertial range relations: the Kolmogorov's celebrated ``four-fifth law'':
\begin{equation}
D_{LLL}(r) = -\frac{4}{5} \epsilon r,
\end{equation}
and a theorem on the velocity-acceleration mixed structure function~\cite{Mann:1999p3829,Pumir:2001p2611,Falkovich:2001p4080,Hill:2006p1631}:
\begin{equation}
\langle\delta_{r} \mathbf{u} \cdot \delta_{r} \mathbf{a}\rangle = -2\epsilon .
\label{eq:duda}
\end{equation}

\begin{figure}
\begin{center}
\psfrag{x}[c][b]{separation $r$ [mm]} 
\psfrag{y}[c][t]{$\epsilon$ [$m^{2}/s^{3}$]} 
  \includegraphics[width=7.8cm]{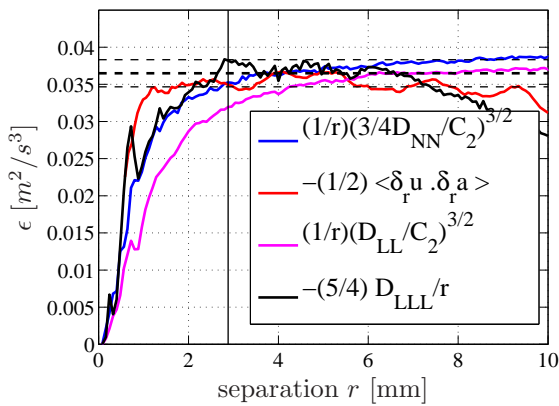}
\end{center}
\caption{\label{fig:epsilon}Illustration of the different methods used to measure the energy dissipation rate $\epsilon$ (see text). 
The horizontal, bold dashed-line shows the measured dissipation $\epsilon = 0.037$ $m^{2}/s^{3}$, the thin dashed-lines represent $\pm 5 \%$ of this value. The vertical line corresponds to the separation $r=40\eta$.}
\end{figure}

We measured $\epsilon$ using each of the four equations above. For length scales in the inertial range ($r/\eta \gtrapprox 50$) we found that measurements from all four methods were in agreement within $\pm 5\%$, as shown in Figure~\ref{fig:epsilon}. We took the weighted average from all four measurements as $\epsilon$.
The integral length scale was estimated as $L = u'^3/\epsilon$, where $u'$ was the fluctuating velocity. Other parameters, such as the Kolmogorov length and time scales, were then obtained from their standard definitions. The turbulence properties and other experimental parameters are summarized in Table~\ref{tab:ExperimentParam}.

\begin{table}
\begin{center}
\begin{tabular}{c c c c c c c}
\hline
\multicolumn{7}{c}{$R_{\lambda}$=442}\\
\hline
 $u'$ & $\epsilon$ & $L$ & $\eta$ & $\tau_{\eta}$ & $N_{f}$ & $\frac{\Delta x}{\eta}$\\
\small{($m/s$)}&\small{($m^{2}/s^{3}$)}&\small{($mm$)}&\small{($\mu m$)}&\small{($ms$)}&\small{(fr$/\tau_{\eta}$)}& \small{(-)}\\
\hline
\hline
$0.15$&$0.036$&$87$&$72$&$5.2$&$31$&$0.6$\\
\hline
\end{tabular}
\end{center}
\caption{Parameters of the experiment. $u'$ is the root-mean-square of the velocity. $\epsilon$ is the  energy dissipation rate per unit mass. $L\equiv u'^{3}/\epsilon$ is the integral length scale. $\eta\equiv (\nu^{3}/\epsilon)^{1/4}$ and $\tau_{\eta}\equiv (\nu/\epsilon)^{1/2}$ are the Kolmogorov length and time scales, respectively, where $\nu$ is the kinematic viscosity of the fluid. $N_{f}$ is the frame rate of the camera, in frames per $\tau_{\eta}$, and $\frac{\Delta x}{\eta}$ is the resolution of the camera compared to $\eta$.}
\label{tab:ExperimentParam}
\end{table}

We now consider the separation of two particles in time $\mathbf{R}(t)\equiv \mathbf{r_{2}}(t)-\mathbf{r_{1}}(t)$ (where $\mathbf{r_{i}}(t)$ stands for the position of particle $i$ time $t$). 
For fluid tracers, it is well known (see e.g.~\cite{falkovich:2001}) that particle velocities are uncorrelated at larger distances ($R \gg L$) and therefore particles separate diffusively ($R \propto t$), while they separate chaotically at separations below Kolmogorov scale ($R \ll \eta$). The interesting question is on the relative dispersion of particles with separations in the inertial range, \textit{i.e.}, $\eta \ll R \ll L$, or equivalently $\eta \ll R_0 \ll L$ and $t \ll T_L$, where $R_{0}\equiv |\mathbf{R}(t=0)|$ is the initial separation and $T_L \equiv L/u' = (L^2/\epsilon)^{1/3}$ is the large eddy turnover time. 
Theoretical work~\cite{richardson:1926,batchelor:1950} suggests the following two regimes for tracer dispersion:
\begin{equation}
\langle\delta\mathbf{R}\cdot\delta\mathbf{R}\rangle = \left \{
\begin{array}{l @{\hspace*{0.5cm}} l}
    \multicolumn{2}{l}{\langle\delta_{r}\mathbf{v}(R_{0})\cdot\delta_{r}\mathbf{v}(R_{0})\rangle t^{2} = \frac{11C_{2}}{3}R_{0}^{2}(\frac{t}{t_{0}})^{2}}\\
    			    & \text{for } t\ll t_{0}=(R_{0}^{2}/\epsilon)^{1/3} \\
  	\multicolumn{2}{l}{\text{and}}\\
    g\epsilon t^{3} & \text{for } t_{0}\ll t\ll T_{L} \\
\end{array}
\right.
\label{separationEquTheorie}
\end{equation}
where $\delta\mathbf{R}(t)\equiv\mathbf{R}(t)-\mathbf{R}(t=0)$ is the vectorial separation increment, $t_{0}$ may be regarded as the life time of an eddy of scale $R_{0}$, and $g$ is a dimensionless coefficient, known as the Richardson constant, expected to be universal and independant of $R_{0}$. While the ballistic regime $\delta\mathbf{R}^{2}\propto t^{2}$ for $t\ll t_{0}$, first predicted by Batchelor~\cite{batchelor:1950}, has been observed numerically \cite{Yeung:2004p1451} and experimentally \cite{Bourgoin:2006p391}, the existence of the Richardson regime ($\delta\mathbf{R}^{2}\propto t^{3}$) is not well established (see \cite{Salazar:2009p2690} for a recent review). 

\begin{figure}
\begin{center}
\psfrag{x}[c][b]{$t/t_{0}$} 
\psfrag{y}[c][t]{$\langle\delta\mathbf{R}\cdot\delta\mathbf{R}\rangle/[\frac{11C_{2}}{3}R_{0}^{2}(\frac{t}{t_{0}})^{2}]$} 
  \includegraphics[width=7.8cm]{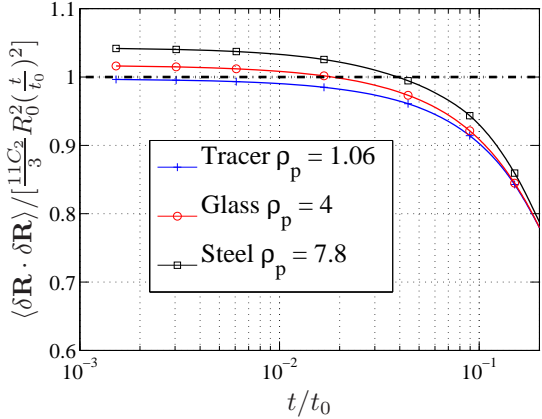}
\end{center}
\caption{\label{separation} Temporal evolution of the relative separation between two particles, for three different types of particle, compared with the Batchelor's law. The initial separation $R_{0}$ is $100\eta$ (within the inertial range).}
\end{figure}

We measured this relative dispersion $\langle\delta\mathbf{R}\cdot\delta\mathbf{R}\rangle (t; R_0)$,  conditioned on the initial separation $R_{0}$, for the three different types of particles. The statistics of inertial particles were expected to be different from that of fluid particles since they do not follow the fluid \cite{MaxeyRiley}. Figure~\ref{separation}  shows the  measured data for a particular initial separation normalized by the Batchelor predictions for tracers.  In agreement with \cite{Bourgoin:2006p391,Ouellette:2006p394}, the fluid particles followed almost perfectly the Batchelor's regime for times smaller than $\sim 0.07t_{0}$. For inertial particles, the $t^{2}$ law still held  for times smaller than $\sim 0.07t_{0}$, but with the prefactor increasing  with particle density. Different initial separations (for constant particle density) showed also Batchelor like scaling, but with different prefactors, as discussed later.
The increase in the prefactor with density reflects  that heavier particles separated initially \emph{faster} than fluid particles. 

\begin{figure}
\begin{center}
\psfrag{x}[c][b]{$r/\eta$} 
\psfrag{y}[c][t]{$\langle\delta_{r}\mathbf{v}\cdot\delta_{r}\mathbf{v}\rangle/[\frac{11C_{2}}{3}(\epsilon r)^{2/3}]$} 
\psfrag{a}[c][b]{$r/\eta$}
\psfrag{b}[c][t]{$\frac{d \ln D_{2}(r)}{d\ln r}$ }  
  \includegraphics[width=7.8cm]{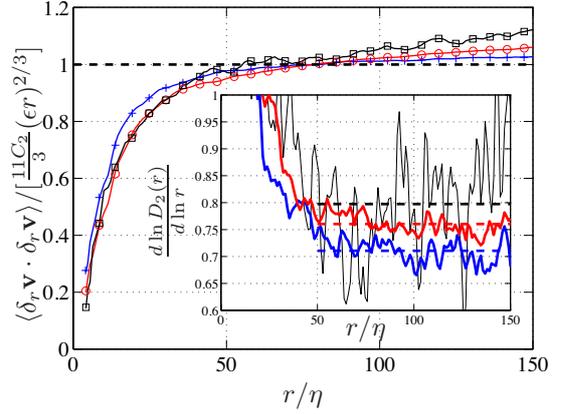}
\end{center}
\caption{\label{deltaVdeltaV}Second order Eulerian structure function $D_{2}(r)=\langle\delta_{r}\mathbf{v}\cdot\delta_{r}\mathbf{v}\rangle$ normalized by $\frac{11C_{2}}{3}(\epsilon r)^{2/3}$ the inertial range value of $D_{2}(r)$ expected for fluid particles. The inset represent the logarithmic slope of $D_{2}(r)$ for the three different types of particles, the horizontal dashed lines represent the average of this quantities for $50<r/\eta<150$. The symbols are the same as the one used in Fig.~\ref{separation}.}
\end{figure}

This last observation, coming from the Lagrangian point of view, is inherently connected to the second order Eulerian structure function. The Batchelor regime was derived from a Taylor expansion of $\langle\delta\mathbf{R}\cdot\delta\mathbf{R}\rangle (t;R_{0}) \approx\langle\delta_{r}\mathbf{v}(R_{0})\cdot\delta_{r}\mathbf{v}(R_{0})\rangle t^{2} \equiv D_{2}(r=R_{0})t^{2}$. As shown in Figure~\ref{deltaVdeltaV}, we measured  the Eulerian quantity $D_{2}(r)$ independently from the particle pair dispersion, which is a Lagrangian measurement. The second order velocity structure function increased systematically when increasing the particle density. This is in agreement with the observation that heavier particles separated faster in the ballistic regime. The structure function $D_{2}(r)$ also increased with length scale $r$ in the inertial range. Therefore the scaling observed and expected for fluid particles $D_{2}\propto (r/\eta)^{2/3}$ applies only approximatively to inertial particles. The inset of Fig.~\ref{deltaVdeltaV} shows the evolution of the logarithmic slope $\frac{d \ln D_{2}(r)}{d \ln r}$ of $D_{2}(r)$ as a function of the scale $r$. The average values of this logarithmic slope, $\xi_{2}$, for $50<r/\eta<150$ are : $\xi_{2}(\text{tracer})=0.71\pm0.02$ (slightly higher than $2/3$, possibly due to small intermittency corrections), $\xi_{2}(\text{glass})=0.76\pm0.02$, and $\xi_{2}(\text{steel})=0.8\pm0.1$ where the errorbars correspond to the standard deviations over the same range of scales.

\begin{figure}
\begin{center}
\psfrag{x}[c][b]{$R_{0}/\eta$} 
\psfrag{y}[c][t]{$\Gamma(R_{0})$} 
  \includegraphics[width=7.8cm]{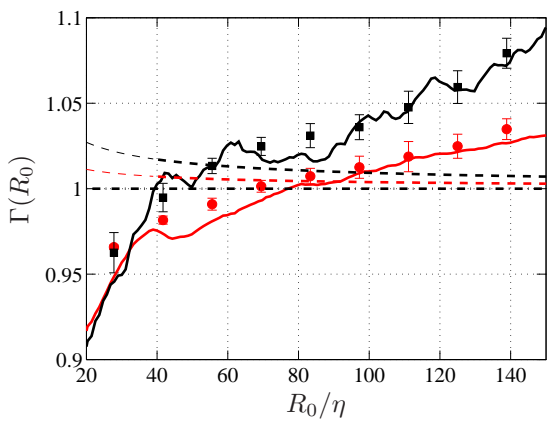}
\end{center}
\caption{\label{RatioDeOuf}Ratio $\Gamma(R_{0})$ (see text). The colors correspond to the two types of inertial particles considered, red for the glass particles ($\rho_p / \rho_f = 4$) and black for the steel particles ($\rho_p / \rho_f = 7.8$). The solid-lines are the Eulerian measurements. Symbols with errorbars are the Lagrangian measurements. The dashed-lines correspond to the model developed in the text.}
\end{figure}

To quantify the effect of inertia on relative dispersion of  particle pairs with initial separations in the inertial range, we define the following ratio:
\begin{equation}
\Gamma (R_{0}) = \frac{\overline{\langle\delta\mathbf{R}\cdot\delta\mathbf{R}\rangle (t,R_{0})}^{t\in [0;0.07t_{0}]}_{inertial}}{\overline{\langle\delta\mathbf{R}\cdot\delta\mathbf{R}\rangle (t,R_{0})}^{t\in [0;0.07t_{0}]}_{fluid}} = \frac{D_{2}(R_{0})_{inertial}}{D_{2}(R_{0})_{fluid}}
\label{EquationGamma}
\end{equation}
where $\overline{X(t)}^{t\in [t_{1};t_{2}]}$ corresponds to a time average of $X(t)$ for $t\in [t_{1};t_{2}]$. This quantity $\Gamma$ could be viewed as a coefficient appearing in the Batchelor regime for inertial particles: $ \langle\delta\mathbf{R}\cdot\delta\mathbf{R}\rangle = \frac{11C_{2}}{3}\Gamma(R_{0})R_{0}^{2}(\frac{t}{t_{0}})^{2} $. 
Note that $\Gamma > 1$ means that the separation velocities between particles are larger than that for fluid tracers.
This ratio, obtained separately from the Lagrangian measurement of relative dispersion and from Eulerian measurement of the second order velocity structure functions, is shown in Figure~\ref{RatioDeOuf}. In the inertial range ($R_{0} \gtrapprox 50\eta$) $\Gamma$ is above one and increases with scale, which confirms that in this range of scales, the inertial particles separated faster than the fluid particles. Moreover, the independent Eulerian and Lagrangian measurements of $\Gamma$ coincide almost perfectly. In the experimentally accessible range of inertial scales $50<R_{0}/\eta<150$, as shown in Fig.~\ref{RatioDeOuf}, $\Gamma$ is mostly greater than unity, and is increasing with $R_{0}$. This reflects that heavy particles separate faster than tracer particles and that the scaling of the second order velocity structure function is affected by the particle inertia. In their numerical simulations,  Salazar \& Collins \cite{SalazarCollins:2010} found a similar effect even if their inertial range is much smaller because of a small Reynolds number ($R_{\lambda}=120$).\\
At sufficiently large separations (on the order of the integral scale $L\sim 10^3\eta$), the velocity spatial correlation vanishes, therefore $\Gamma$ tends to the ratio of the average particle kinetic energy per unit mass $k_{p}=\frac{1}{2}\langle\mathbf{v}\cdot\mathbf{v}\rangle$ to that of the fluid $k_{f}=\frac{1}{2}\langle\mathbf{u}\cdot\mathbf{u}\rangle$. In our measurements, surprisingly, this ratio is bigger than unity and varies non-monotonically with increasing particle density $\rho_{p}$ : $k_{p}(\rho_{p}=4)/k_{f}=1.063$ and $k_{p}(\rho_{p}=7.8)/k_{f}=1.05$. In a recent numerical work, Salazar \& Collins \cite{SalazarCollins:2010} observed the same effect using a \textit{point-particle model}: With increasing $St$, this ratio first increased to $St\lessapprox 0.25$ and then fell  bellow unity. Moreover, by computing the fluid kinetic energy along the inertial particles trajectories they demonstrated clearly, that for $St<0.5$ the non-monotonic behaviour of $k_{p}$ with $St$ is due to the uneven sampling of the flow by inertial particles.

As an attempt to understand analytically the observed increase of relative velocity with particle inertia, we can start  with the simplest equation of motion for the inertial particles:
\begin{equation}
\frac{d\mathbf{v}}{dt}=\beta \frac{D\mathbf{u}}{Dt}+\frac{1}{\tau_{p}}(\mathbf{u}-\mathbf{v})+(1-\beta)\mathbf{g}, 
\label{EquMouv}
\end{equation}
where $\mathbf{v}$ and $\mathbf{u}$ are the velocities of the particles and the fluid at the particle position, 
the modified density ratio $\beta \equiv 3\rho_{f}/(2\rho_{p}+\rho_{f})$ takes into account the added mass effect, 
$\tau_{p}\equiv d_{p}^{2}/12\beta\nu$ is the particle viscous relaxation time, 
$\mathbf{g}$ is the gravity vector and the operators $d/dt$ and $D/Dt$ indicate derivatives following an inertial particle or a fluid element, respectively.
Eq.~\eqref{EquMouv} is the simplest form of the equation derived for inertial particles (see, e.g.~\cite{MaxeyRiley,GATIGNOL:1983p3869}) and has been used before to study the effect of particle inertia~\cite{Falkovich:2004p987}.
Using this equation, one can solve for particle velocity in the limit of small $\tau_{p}$ (\textit{i.e.}, $St \ll 1$):
\begin{equation}
\begin{array}{l @{=} l}
    \mathbf{v} & \mathbf{u}+ \tau_{p}(1-\beta)(\mathbf{g}-\mathbf{a})+o(\tau_{p}^{2}) \\
\end{array}
\label{EquVitessAcc}
\end{equation}
With this we  evaluate the Eulerian structure function of the inertial particles in terms of statistical quantities of the flow field:
\begin{equation}
\begin{array}{l @{} l}
     \langle\delta_{r}\mathbf{v}\cdot\delta_{r}\mathbf{v}\rangle &  =\langle\delta_{r}\mathbf{u}\cdot\delta_{r}\mathbf{u}\rangle+2(\beta-1)\tau_{p}\langle\delta_{r}\mathbf{u}\cdot\delta_{r}\mathbf{a}\rangle+o(\tau_{p}^{2}),
\end{array}
\label{EulerianStructFunc}
\end{equation}
where $\mathbf{a} = D\mathbf{u}/Dt$ is the fluid acceleration.
We can compare directly the results of this model to the measurements by computing $\Gamma(r)$ (see equation \ref{EquationGamma}):
\begin{equation}
\Gamma(r) = \frac{\langle\delta_{r}\mathbf{v}\cdot\delta_{r}\mathbf{v}\rangle}{\langle\delta_{r}\mathbf{u}\cdot\delta_{r}\mathbf{u}\rangle} =1+ \frac{12}{11C_{2}}(1-\beta)St\left(\frac{r}{\eta}\right)^{-2/3} .
\label{EquationGammaTheor}
\end{equation}
This expression is only valid for length scales in the inertial range ($r \gtrapprox 50\eta$) since we used the relation $\langle\delta_{r}\mathbf{u}\cdot\delta_{r}\mathbf{a}\rangle = -2\epsilon$ as given in Eq.~\eqref{eq:duda}. 
For heavy particles ($\rho_p > \rho_f$), the density ratio $\beta < 1$. Our simple equation predicts that $\Gamma > 1$, \textit{i.e.}, heavier particles possess larger relative velocities and hence separate faster.
This is in qualitative agreement with the experimental observation. However, as shown in Figure~\ref{RatioDeOuf}, the quantitative agreement between the simple model and the measurements is not satisfactory. In particular, the simple model shows that the relative increase of relative velocity between inertial particles is proportional to a ``scale-dependent Stokes number'' $St_{r}=St(\eta / r)^{2/3} = \tau_{p}/\tau_{r}$, which compares the particle time scale to the life time of an eddy of scale $r$, $\tau_{r}=(r^{2}/\epsilon)^{1/3}$ and hence \emph{decreases} with scale $r$, while the measurements shows that effect \emph{increases} with scale in the range accessible experimentally.

We now discuss the reasons why this simple model does not agree quantitatively with the measurements. At first sight, it might come by neglecting  terms of order $\tau_{p}^{2}$ and higher. However, as one can see from Figure~\ref{RatioDeOuf}, if this was the case, it would mean that the higher order terms are actually the leading terms, which is unlikely. The second possible explanation comes from the fact that in the last step of Eq.~\eqref{EquationGammaTheor}, we implicitly assumed that inertial particles distribute evenly in space. This assumption ignored the so called "sampling effect" of inertial particles, as it has been shown that heavy particles are ejected by intense vortices and tend to accumulate in low-vorticity/high-strain regions~\cite{sundaram:1997,Bec:2006p656}. Therefore, these heavy particles might experience a higher "dissipation rate". As a first order correction to this effect, we replaced the term $\langle \delta_r{\mathbf{u}} \cdot \delta_r{\mathbf{a}}\rangle$ in Eq.~\eqref{EulerianStructFunc} by the measured value of $\langle \delta_r{\mathbf{v}} \cdot \delta_r{\mathbf{a_p}}\rangle$, where $\mathbf{a_p}$ was the acceleration of inertial particles. The calculated $\Gamma(r)$ still does not agree with measurements, especially its dependence on scale $r$. 
Using numerical simulations, Salazar \& Collins \cite{SalazarCollins:2010} showed that Eq.~\ref{EulerianStructFunc} led to $\Gamma(r)$ that agrees reasonably well with their DNS if the quantities on the right hand side of Eq.~\ref{EulerianStructFunc} were conditioned on the particle trajectories. In order to further identify, whether sampling or  filtering \cite{SalazarCollins:2010} is responsible for the apparent discrepancy between the model and the data, we estimated  with Eq. \ref{EquVitessAcc} the fluid velocity at the particle position , up to order $\tau_p^2$ (using $\mathbf{a_{p}}=\mathbf{a}+o(\tau_{p})$).  For all three Stokes numbers the second order velocity structure functions of this estimated fluid velocity field showed the same logarithmic slopes as those of the measured particle velocity field. Thus, the lack of agreement between the experimental data and the  simple model, Eq.~\ref{EquationGammaTheor}, most likely roots in the uneven sampling of the flow field by inertial particles and not in the filtering of the velocity field by inertial particles. At the Stokes numbers observed the non-differentiable particle velocity field (caustics) can also contribute as discussed by Salazar \& Collins \cite{SalazarCollins:2010}, however, we believe this to be a less important effect.\\

We summarize our experimental results as follows. 
For heavy, Kolmogorov-sized particles we have observed a ballistic, or Batchelor regime for times $t \lessapprox 0.1 t_0$ with $t_0$ determined by the turbulence energy dissipation rate and the initial separation between particle pairs.  In this regime, these heavy, inertial particles separated faster than fluid tracers, \textit{i.e.}, the relative velocity between particles increased with particle inertia. This was also reflected by the  measured Eulerian velocity structure functions. In the inertial range, the logarithmic slope of the second order velocity structure function increased with the particle density. We also observed a non-monotonic variation of the RMS velocity  with Stokes number, which, in this range of Stokes numbers, Salazar \& Collins \cite{SalazarCollins:2010} attributed to the preferential sampling of the inertial particles.
The observed effect of inertia on particle dynamics could not be captured by a model based on the simplified equation of motion for idealized point-like particles. The inaccuracy  of the model may originate from the simplified model assumptions that do not capture the preferential sampling of inertial particles \cite{PRLus:2010}. In addition, a point-particle model might not be sufficient to describe the dynamics of the Kolmogorv size particles studied here.  Finally we note that the increase of relative velocity between inertial particles with their density will result in an increase of the collision rate between inertial particles, which  has important consequences in problems such as rain formation in warm clouds. To understand  these issues quantitatively, we would need to spatially resolve the flow at the particle scale. This is currently under investigation.\\

\acknowledgments
We acknowledge J. Bec, M. Bourgoin, R. J. Hill, J.-F. Pinton and A. Pumir for many interesting discussions. This work was funded by the Max Planck Society, and the Marie Curie Fellowship, Programme PEOPLE - Call FP7-PEOPLE-IEF-2008  Proposal N$^{o}$ 237521. 
\bibliographystyle{eplbib}
\bibliography{EPL_particule}

\begin{thebibliography}{10}
\expandafter\ifx\csname url\endcsname\relax\def\url#1{\texttt{#1}}\fi

\bibitem{seminara:2010}
\Name{Seminara G.} \REVIEW{Ann. Rev. Fluid Mech. }{42}{2010}{43}.

\bibitem{Schmitt:2008p2887}
\Name{Schmitt F.~G. \and Seuront L.} \REVIEW{J. Marine. Syst. }{70}{2008}{263}.

\bibitem{denman:1995}
\Name{Denman K.~L. \and Gargett A.~E.} \REVIEW{Ann. Rev. Fluid Mech.
  }{27}{1995}{225}.

\bibitem{Meiburg2009rev}
\Name{Meiburg E. \and Kneller B.} \REVIEW{Ann. Rev. Fluid Mech.
  }{42}{2010}{135}.

\bibitem{Lewellen:2008p2927}
\Name{Lewellen D.~C., Gong B. \and Lewellen W.~S.} \REVIEW{J. Atmos. Sci.
  }{65}{2008}{3247}.

\bibitem{Waller:2008p3195}
\Name{Waller D., Greeley R., Neakrase L.~D., Sullivan R., Johnson J. \and Team
  A.~S.} \REVIEW{39th Lunar and Planetary Science Conference }{39}{2008}{2218}.

\bibitem{Shaw:2003p769}
\Name{Shaw R.~A.} \REVIEW{Ann. Rev. Fluid Mech. }{35}{2003}{183}.

\bibitem{Klahr:2006p3103}
\Name{Klahr H. \and Brandner W.} (Editors) \Book{Planet Formation: Theory,
  Observations and Experiments} (Cambridge Univ. Press, Cambridge, UK) 2006.

\bibitem{Bec:2006p656}
\Name{Bec J., Biferale L., Boffetta G., Celani A., Cencini M., Lanotte A.,
  Musacchio S. \and Toschi F.} \REVIEW{J. Fluid Mech. }{550}{2006}{349}.

\bibitem{Ferrante:2003p3145}
\Name{Ferrante A. \and Elghobashi S.} \REVIEW{Phys. Fluids }{15}{2003}{315}.

\bibitem{Bec:submitted}
\Name{Bec J., Biferale L., Lanotte A.~S., Scagliarini A. \and Toschi F.}
  \REVIEW{J. Fluid Mech. }{645}{2010}{497}.

\bibitem{Zaichik:2009p3167}
\Name{Zaichik L.~I., Simonin O. \and Alipchenkov V.~M.} \REVIEW{J. Turbul.
  }{10}{2009}{4}.

\bibitem{snyder:1971}
\Name{Snyder W. \and Lumley J.} \REVIEW{J. Fluid Mech. }{48}{1971}{41}.

\bibitem{Ayyalasomayajula:2006p267}
\Name{Ayyalasomayajula S., Gylfason A., Collins L.~R., Bodenschatz E. \and
  Warhaft Z.} \REVIEW{Phys. Rev. Lett. }{97}{2006}{144507}.

\bibitem{Qureshi:2007p1316}
\Name{Qureshi N., Bourgoin M., Baudet C. \and Cartellier A.} \REVIEW{Phys. Rev.
  Lett. }{99}{2007}{184502}.

\bibitem{Xu:2008p1025}
\Name{Xu H. \and Bodenschatz E.} \REVIEW{Physica D }{237}{2008}{2095}.

\bibitem{Volk:2008p271}
\Name{Volk R., Mordant N., Verhille G. \and Pinton J.~F.} \REVIEW{Europhys
  Lett. }{81}{2008}{34002}.

\bibitem{MaxeyRiley}
\Name{Maxey M. \and Riley J.} \REVIEW{Phys. Fluids }{26}{1983}{883}.

\bibitem{GATIGNOL:1983p3869}
\Name{Gatignol R.} \REVIEW{J Mec Theor Appl }{2}{1983}{143}.

\bibitem{PartRev09}
\Name{Toschi F. \and Bodenschatz E.} \REVIEW{Ann. Rev. Fluid Mech.
  }{41}{2009}{375}.

\bibitem{Bourgoin:2006p391}
\Name{Bourgoin M., Ouellette N.~T., Xu H., Berg J. \and Bodenschatz E.}
  \REVIEW{Science }{311}{2006}{835}.

\bibitem{Ouellette:2006p384}
\Name{Ouellette N.~T., Xu H. \and Bodenschatz E.} \REVIEW{Exp. Fluids
  }{40}{2006}{301}.

\bibitem{Voth:2002p432}
\Name{Voth G.~A., {La Porta} A., Crawford A.~M., Alexander J. \and Bodenschatz
  E.} \REVIEW{J. Fluid Mech. }{469}{2002}{121}.

\bibitem{Xu:2008p1921}
\Name{Xu H.} \REVIEW{Meas. Sci. Technol. }{19}{2008}{075105}.

\bibitem{Mordant:2004p245}
\Name{Mordant N., Crawford A.~M. \and Bodenschatz E.} \REVIEW{Physica D
  }{193}{2004}{245}.

\bibitem{sreenivasan:1995}
\Name{Sreenivasan K.~R.} \REVIEW{Phys. Fluids }{7}{1995}{2778}.

\bibitem{Mann:1999p3829}
\Name{Mann J., Ott S. \and Andersen J.} \REVIEW{Ris{\o}--R--1036(EN)
  }{}{1999}{}.

\bibitem{Pumir:2001p2611}
\Name{Pumir A., Shraiman B. \and Chertkov M.} \REVIEW{Europhys. Lett.
  }{56}{2001}{379}.

\bibitem{Falkovich:2001p4080}
\Name{Falkovich G., Gaw{\c e}dzki K. \and Vergassola M.} \REVIEW{Rev. Mod.
  Phys. }{73}{2001}{913}.

\bibitem{Hill:2006p1631}
\Name{Hill R.} \REVIEW{J Turbul }{7}{2006}{1}.

\bibitem{falkovich:2001}
\Name{Falkovich G., Gawedzki K. \and Vergassola M.} \REVIEW{Rev. Mod. Phys.
  }{73}{2001}{913}.

\bibitem{richardson:1926}
\Name{Richardson L.~F.} \REVIEW{Proc. R. Soc. Lond. A }{110}{1926}{709}.

\bibitem{batchelor:1950}
\Name{Batchelor G.~K.} \REVIEW{Q. J. R. Meteor. Soc. }{76}{1950}{133}.

\bibitem{Yeung:2004p1451}
\Name{Yeung P. \and Borgas M.} \REVIEW{J. Fluid Mech. }{503}{2004}{93}.

\bibitem{Salazar:2009p2690}
\Name{Salazar J. P. L.~C. \and Collins L.~R.} \REVIEW{Ann. Rev. Fluid Mech.
  }{41}{2009}{405}.

\bibitem{Ouellette:2006p394}
\Name{Ouellette N.~T., Xu H., Bourgoin M. \and Bodenschatz E.} \REVIEW{New
  Journal of Physics }{8}{2006}{109}.

\bibitem{SalazarCollins:2010}
\Name{Salazar J. P. L.~C. \and Collins L.~R.} \REVIEW{Submitted to J. Fluid
  Mech. }{-}{2010}{}.

\bibitem{Falkovich:2004p987}
\Name{Falkovich G. \and Pumir A.} \REVIEW{Phys. Fluids }{16}{2004}{L47}.

\bibitem{sundaram:1997}
\Name{Sundaram S. \and Collins L.~R.} \REVIEW{J. Fluid Mech. }{335}{1997}{75}.

\bibitem{PRLus:2010}
\Name{Gibert M., Xu H. \and Bodenschatz E.} \REVIEW{under revision, Phys. Rev.
  Lett., arXiv:1002.3755 }{-}{2010}{}.

\end{thebibliography}
\end{document}